\documentclass[%
reprint,
superscriptaddress,
amsmath,amssymb,
aps,
pra,
twocolumn
]{revtex4-2}
\usepackage{makecell}
\usepackage{graphicx}
\usepackage{dcolumn}
\usepackage{bm}
\usepackage{amsmath}
\usepackage{booktabs}
\DeclareMathOperator{\Tr}{Tr}
\usepackage{xcolor}

\usepackage{comment}
\usepackage{hyperref}

\begin{document}

\title{
Gravitational Decoherence Estimation in Optomechanical Systems}
\author{Leonardo A. M. Souza}
 \affiliation{Universidade Federal de Viçosa - Campus Florestal, LMG 818 Km6 S/N, Florestal, 35690-000, Minas Gerais, Brazil.}
\author{Olimpio P. de S\'a Neto}%
 \email{olimpiopereira@phb.uespi.br}
\affiliation{ Coordena\c{c}\~ao de Ci\^encia da Computa\c{c}\~ao, Universidade Estadual do Piau\'i, 64202-220, Parna\'iba (PI), Brazil. }
\author{Enrico Russo}
\affiliation{Dipartimento di Ingegneria, Università degli Studi di Palermo, Viale delle Scienze, 90128
Palermo, Italy}
\affiliation{University San Pablo-CEU, CEU Universities, Department of Applied Mathematics and Data Science, Campus de Moncloa, C/Juli\'{a}n Romea 23 28003, Madrid, Spain}
\author{Rosario Lo Franco}
\affiliation{Dipartimento di Ingegneria, Università degli Studi di Palermo, Viale delle Scienze, 90128
Palermo, Italy}
\author{Gerardo Adesso}
\affiliation{School of Mathematical Sciences and Centre for the Mathematics and Theoretical Physics of Quantum Non-Equilibrium Systems, University of Nottingham, University Park, Nottingham NG7 2RD, UK}




\date{\today}

\begin{abstract}
We develop a comprehensive quantum estimation framework to quantify how precisely gravitationally induced decoherence can be inferred in optomechanical systems, using single-mode Gaussian probe states. Our approach combines a microscopic description of the gravitational diffusion mechanism with quantum Fisher information to determine the ultimate sensitivity achievable in principle. We show that gravitational diffusion leaves distinct, measurable signatures in the mechanical state, both during transient evolution and in the stationary regime. Finally, we identify how probe state preparation shapes the attainable precision, thereby establishing fundamental limits for detecting and estimating gravity-driven decoherence.


\end{abstract}

\maketitle

\section{Introduction}

The detection of gravitational waves has rapidly transitioned from a rare achievement to a routinely observed phenomenon \cite{abbott2016observation, tse2019quantum}, revolutionizing the landscape of observational astrophysics. This progress has been made possible by the collaborative efforts of leading research groups \cite{abbott2016observation, aasi2015advanced}, whose innovations in quantum-enhanced instrumentation have significantly improved the sensitivity of gravitational-wave observatories. In particular, the incorporation of quantum optical techniques has allowed for the mitigation of quantum noise, an intrinsic limitation in such detectors \cite{tse2019quantum, abbott2020gw190412}, thereby extending the observable reach of these facilities \cite{aasi2015advanced}. 

Beyond the pioneering work of LIGO and Virgo, a global network of gravitational-wave observatories has emerged, each contributing unique capabilities to the field \cite{milton2011hard}. The KAGRA detector in Japan combines underground operation with cryogenic mirrors to reduce seismic and thermal noise \cite{akutsu2018construction}, while GEO600 in Germany has served as a tested for advanced quantum techniques, including squeezed light injection and signal recycling \cite{affeldt2014advanced}. Looking forward, the Einstein Telescope in Europe and the Cosmic Explorer in the United States are proposed next-generation observatories that promise order-of-magnitude improvements in sensitivity. Additionally, space-based missions such as LISA (Laser Interferometer Space Antenna) aim to detect lower-frequency gravitational waves from supermassive black hole mergers and cosmological sources \cite{ligo2025gw231123,omonster}, extending the reach of gravitational-wave astronomy to entirely new regimes \cite{hardwick2019high,vinet2009special,harry2010advanced}.

\begin{figure}[b!]
    \centering
    \includegraphics[width=1\columnwidth]{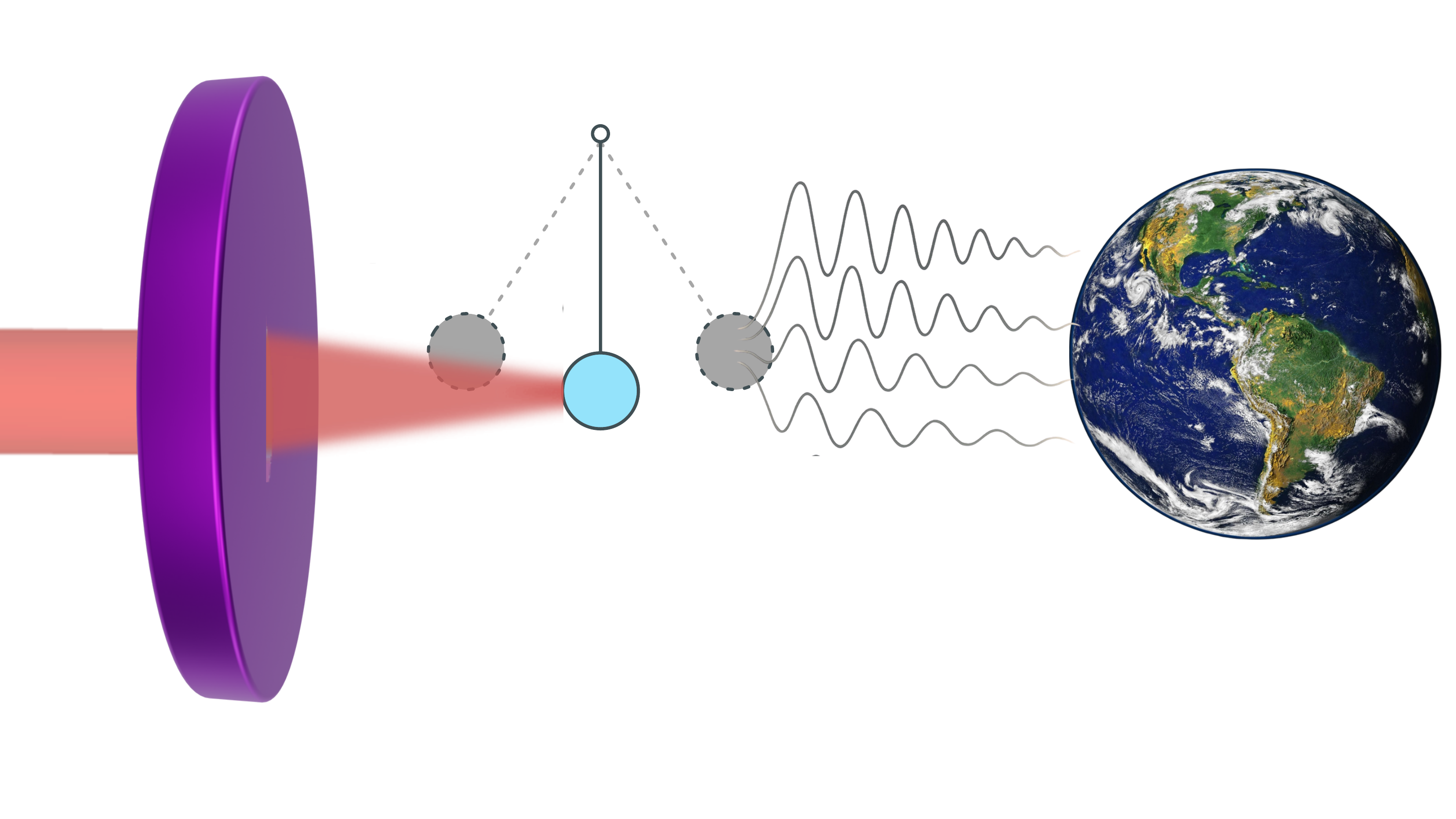}
    \caption{{\bf A pictorical scheme of the phenomenon:} Conceptual diagram of an optomechanical system subject to gravitational decoherence. A mechanical oscillator (a simple pendulum as a moving mirror) coupled to an optical field in a cavity is used as a quantum probe. }
    \label{gravity}
\end{figure}

Gravitational-wave interferometers rely on laser interferometry: coherent light is split into two beams that traverse orthogonal, kilometer-scale arms and then recombine, so that differential phase shifts reveal extremely small perturbations of space-time. Their sensitivity, however, is ultimately constrained by the quantum statistics of light. In particular, photon-number fluctuations give rise to shot noise, which manifests as uncertainty in the detected phase (or equivalently in the effective arrival time of photons) and produces a Gaussian spread in the measurement outcomes around the mean. A signal becomes resolvable when the gravitational-wave–induced change in the optical time of flight (or phase) exceeds the characteristic width of this quantum-noise distribution \cite{abbott2016gw151226}.


To overcome this limitation, the LIGO and Virgo collaborations have implemented the quantum squeezing technique \cite{ligomemorandum}. Originally proposed by Carlton Caves \cite{caves1980quantum}, this method effectively narrows the temporal distribution of photon arrivals, thereby reducing the impact of vacuum fluctuations on the signal. Notably, squeezing is particularly beneficial at higher frequencies, enabling improved localization of gravitational-wave sources in the sky. This, in turn, allows to correlate gravitational signals with electromagnetic counterparts, that led to developments in multi-messenger astronomy \cite{hardwick2019high}. The implications are profound: events such as black hole mergers, neutron star collisions, and possibly even supernovae, are now more readily accessible to observation with heightened precision \cite{latune2013quantum}.

While gravitational-wave detectors exemplify the forefront of this endeavour, atomic interferometry also plays a crucial role in modern physics. Applications range from geophysical surveys \cite{farah2014underground}, inertial navigation \cite{geiger2011detecting}, and geoid mapping \cite{novak2003geoid}, to fundamental tests of general relativity \cite{imanishi2004network,geneves2005bnm}. These systems provide a compelling platform for precision experiments probing the Einstein equivalence principle, particularly regarding whether gravitational acceleration depends on internal quantum states.

In light of these developments, we turn our attention to {the estimation of gravitationally induced decoherence rate} within gravitational-wave-inspired frameworks. Adopting the formalism of quantum Fisher information \cite{fisher1925theory,giovannetti2011advances},
{we look for states of the probe that allow to attain the ultimate precision limit dictated by quantum metrology} \cite{aspachs2010optimal,piotrowski2023simultaneous}. Building on prior methodologies \cite{helstrom1976quantum}, we propose and compare quantum detection schemes aimed at estimating parameters, including temperature, through numerical simulations \cite{latune2013quantum,holevo2011probabilistic,helstrom1969quantum,helstrom2013statistical,de2022temperature,de2022simple,marin2013gravitational,amati1987superstring,vasileiou2015planck,chung2021upper,harney2021ultimate,spedalieri2018thermal,rubio2021global,alves2022bayesian,planella2022bath,luiz2022machine,cavina2018bridging,correa2015individual,mehboudi2019thermometry,chowdhury2019calibrated,pati2020quantum,seveso2018trade,hofer2019fundamental,abiuso2020optimal,correa2017enhancement,de2019estimation, Leo2024}. Furthermore, we {put forward} how optomechanical platforms \cite{qvarfort2018gravimetry,sala2021quantum}, via radiation pressure coupling between mechanical oscillators and cavity fields, offer an experimentally viable route to explore gravitational decoherence (see Fig.~\ref{gravity} for an illustration of the phenomenon). Finally, we present a discussion on the implications of this study and conclude our presentation with the obtained results.

\section{Physical systems}

\subsection{Gaussian states}\label{gauss_sec}

In this work we are interested in single-mode Gaussian states subjected to Gaussian dynamics \cite{adesso2007entanglement, adesso2007entanglementThesis, adesso2014continuous, cerf2007quantum, serafini2005quantifying, Leo2024}. A continuous variable system of one mode can be defined by the quadrature vector $\hat{\boldsymbol{O}} = ( \hat{x}, \hat{p} )$, where 
$\hat{x} = \sqrt{\frac{\hbar}{2m\omega_m}} \left( \hat{b} + \hat{b}^\dagger \right)$ and $\hat{p} = -i \sqrt{\frac{\hbar m \omega_m}{2}} \left( \hat{b} - \hat{b}^\dagger \right)$, with $\hat{b}$ and $\hat{b}^\dagger$ the bosonic annihilation and creation operators, respectively. The quadratures obey the canonical commutation relations $ [\hat{O}_j, \hat{O}_k] = i \Omega_{jk}$, with $\boldsymbol{\Omega}=(\Omega_{jk})$ being the symplectic form $\boldsymbol{\Omega} =
\left(
\begin{array}{cc}
0 & 1 \\                                  -1 & 0 \\                                 \end{array}                               \right).$ A Gaussian state $\rho$ is represented by a Gaussian characteristic function in phase space
, and is completely characterized by its first and second statistical moments of the quadrature vector, given respectively by the displacement vector $\boldsymbol{\varepsilon} = (\varepsilon_j)$ and the covariance matrix  $\boldsymbol{\sigma} = (\sigma_{jk})$, where $\varepsilon_j = \langle \hat{O}_j \rangle$ and $\sigma_{jk} = \frac{1}{2} \langle \hat{O}_j \hat{O}_k + \hat{O}_k \hat{O}_j \rangle - \langle \hat{O}_j \rangle \langle \hat{O}_k \rangle$. A \textit{bona fide} condition satisfied by all physical Gaussian states is the Robertson-Schr\"{o}dinger uncertainty relation, given by
\begin{equation}\label{bonafide}
\boldsymbol{\sigma} + \frac{i}{2} \boldsymbol \Omega \geq 0.
\end{equation}

For our purpose (since we deal with single-mode Gaussian states), a general covariance matrix can be written as \cite{Luca} 
\begin{equation}
\boldsymbol{\sigma} = \left(              \begin{array}{cc}                       a_1 & g \\
g & b_1
\end{array}                               \right),
\label{CMgeneral1}
\end{equation}
where the coefficients are defined such that (\ref{CMgeneral1}) satisfies the physical constraint (\ref{bonafide}). The mean number of excitations (proportional to the total mean energy) of a single-mode Gaussian state can be defined as: $E \equiv {n_0},$ where ${n_0} = (\Tr [\boldsymbol{\sigma}] - 1)/2 + (\varepsilon_{x}^2 + \varepsilon_{p}^2)/2$ 
\cite{adesso2007entanglementThesis, Leo2024, Luca}.

\subsection{Quantum Estimation Theory}

The goal of any estimation strategy, classical or quantum, is to extract information about a parameter encoded in a system with maximum precision. This typically involves encoding the parameter in $N$ copies of the system, followed by an optimal measurement to infer its value. Estimation is thus based on statistical theory \cite{helstrom1976quantum, holevo2003statistical, ballester2005estimation, paris2009quantum}.

The goal is to construct an unbiased estimator $\hat{\epsilon}$ for a parameter $\epsilon$ by sampling a random variable $x$ drawn from a distribution $p(x|\epsilon)$ that depends on $\epsilon$. For a measurement strategy $M$, the Cramér-Rao bound sets a lower limit on the variance of any unbiased estimator: $\text{VAR}(\hat{\epsilon}) \geq 1/(N F_\epsilon^M)$, where $F_\epsilon^M$ is the classical Fisher information. In the quantum case, optimizing over all possible measurements leads to the quantum Cramér-Rao bound: $\text{VAR}(\hat{\epsilon}) \geq 1/(N \mathcal{F}_\epsilon)$, where $\mathcal{F}_\epsilon$ is the Quantum Fisher Information (QFI), defined via the symmetric logarithmic derivative (SLD) $L_\epsilon$ as 
\begin{equation}\mathcal{F}_\epsilon = \Tr{\rho_\epsilon L_\epsilon^2}, \mbox{\quad with \quad} \rho_\epsilon L_\epsilon + L_\epsilon \rho_\epsilon = 2\, \partial_\epsilon \rho_\epsilon.
\end{equation}

Two key points are worth noting: (i) the Fisher Information is generally bounded by the Quantum Fisher Information: $F_\epsilon^M \leq \mathcal{F}_\epsilon$; (ii) using quantum resources, such as entanglement, squeezing, or Fock states, one can surpass the classical shot noise limit and approach the Heisenberg limit, potentially yielding a quadratic improvement in estimation precision \cite{adesso2009optimal, giovannetti2011advances, giovannetti2006quantum,matsumura2020gravity}. In this work, while entangled Gaussian states are not considered, squeezing is still anticipated to improve the performance of the estimation strategy analysed.


\subsection{Dynamics}

To gain a deeper understanding of the significance of gravitational decoherence, we examine Diosi's theory of gravitational decoherence \cite{diosi1989models} as an example. This is equivalent to the decoherence model introduced by Kafri et al \cite{kafri2014classical}. Diosi’s model can be applied to an optomechanical cavity where one mirror is free to move within a harmonic potential with a specific frequency. The master equation for a massive particle moving in a harmonic potential, considering gravitational decoherence, is given by 
\begin{eqnarray} 
\frac{d\hat{\rho}}{dt} &=& -i \omega_{m} \left[ \hat{b}^{\dag}\hat{b},\hat{\rho} \right] - \Lambda \left[ \hat{b}+\hat{b}^{\dag}, \left[ \hat{b}+\hat{b}^{\dag} , \hat{\rho} \right] \right]\nonumber\\
&& + \gamma (\bar{n}_{th} + 1) \left( \hat{b} \hat{\rho} \hat{b}^{\dag} - \frac{1}{2} \left\{ \hat{b}^{\dag} \hat{b}, \hat{\rho} \right\} \right) \nonumber\\
&& + \gamma \bar{n}_{th} \left( \hat{b}^{\dag} \hat{\rho} \hat{b} - \frac{1}{2} \left\{ \hat{b} \hat{b}^{\dag}, \hat{\rho} \right\} \right), 
\label{masterequation} 
\end{eqnarray} 
where $\hat{b} = \sqrt{\frac{m\omega_{m}}{2\hbar}}\hat{x} + i \frac{1}{\sqrt{2\hbar m\omega_{m}}}\hat{p}$,
with $\hat{x}$ and $\hat{p}$ being the usual canonical position and momentum operators for the moving mirror, $\bar{n}_{th}$ is thermal occupation number, $\gamma$ is the thermal dissipation rate. The parameter $\Lambda$ is a decoherence noise rate 
\begin{equation} 
\Lambda = \Lambda_{g} + \Lambda_{T}, 
\end{equation} 
{that includes both gravitational and thermal decoherence effects, respectively}:
\begin{equation} \Lambda_{g} = \frac{2\pi G\Delta}{3\omega_{m}}, \quad \Lambda_{T} = \frac{k_{B}T \gamma}{\hbar \omega_m},\end{equation}
with $G$ representing Newton’s gravitational constant, $\Delta$ the density of a moving mirror, $k_{B}$ and $\hbar$ the Boltzmann and Planck constants, respectively. 
As expected, $\Lambda_{g}$ is quite small, of the order of $10^{-8}\ \mathrm{s}^{-1}$ for suspended mirrors with $\omega_{m}\sim 1 \ \mathrm{s}^{-1}$. It is clear that, to highlight the gravitational term’s effect in contrast to mechanical heating, the temperature $T$ has to be kept low.

From Eq.~(\ref{masterequation}), we have the equations of motion
    \begin{eqnarray}
        \frac{d}{dt}  \big\langle \hat{b} \big\rangle &=& - \left( i\omega_{m}  + \frac{\gamma}{2} \right) \big\langle \hat{b} \big\rangle , \\
        \frac{d}{dt} \big\langle \hat{b}^{\dagger}\hat{b} \big\rangle &=& -  \gamma \big\langle \hat{b}^{\dagger}\hat{b} \big\rangle  + \gamma \bar{n}_\mathrm{eff}, \\
        \frac{d}{dt} \big\langle \hat{b}^{2} \big\rangle &=& - \left( 2i\omega_{m} + \gamma \right) \big\langle \hat{b}^{2} \big\rangle + 2 \Lambda,
    \end{eqnarray}
where we have introduced $\bar{n}_{\text{eff}} = \bar{n}_{\text{th}} + \frac{2 \Lambda}{\gamma}$, that is the effective occupation number which incorporates both the thermal contribution from the environment and an additional noise term proportional to $\Lambda/\gamma$. This shows that position diffusion, {analogously to thermal excitation}, may increase the average energy of the harmonic oscillator and, therefore, raise its effective temperature. The above equations integrate to
    \begin{eqnarray}
        \big\langle \hat{b} \big\rangle_{t} &=& e^{- \left( i\omega_{m}  + \frac{\gamma}{2} \right) t } \big\langle \hat{b} \big\rangle_{0} , \label{bt} \\
        \big\langle \hat{b}^{\dagger}\hat{b} \big\rangle_{t} &=&
         \big\langle \hat{b}^{\dagger}\hat{b} \big\rangle_{0}
         e^{-\gamma t} + \left( 1- e^{-\gamma t}\right) \bar{n}_{eff}, \\
        \big\langle \hat{b}^{2} \big\rangle_{t} &=& e^{- \left( 2i\omega_{m} + \gamma \right) t } \big\langle \hat{b}^{2} \big\rangle_{0} \nonumber \\ && + \frac{ 2 \Lambda}{ \left( 2i\omega_{m} + \gamma \right)} \left[ 1-e^{- \left( 2i\omega_{m} + \gamma \right) t } \right]. \label{b2t}
    \end{eqnarray} 
    The covariance matrix is \cite{adesso2007entanglement, adesso2007entanglementThesis, adesso2014continuous, cerf2007quantum, serafini2005quantifying, Leo2024, nemesnotes}
    \begin{equation}
\label{covariance}
C(\hat b,\hat b^\dagger)=
\begin{pmatrix}
\langle \hat b^{2} \rangle_t - \langle \hat b \rangle_t^{2} &
\langle \hat b^\dagger \hat b \rangle_t + \tfrac12
- \langle \hat b^\dagger \rangle_t \langle \hat b \rangle_t \\
\langle \hat b^\dagger \hat b \rangle_t + \tfrac12
- \langle \hat b^\dagger \rangle_t \langle \hat b \rangle_t &
\langle \hat b^{\dagger 2} \rangle_t - \langle \hat b^\dagger \rangle_t^{2}
\end{pmatrix}.
\end{equation}

The general solution for the dynamics of Gaussian states can be expressed {both} in terms of the bosonic operators $\hat{b}$ and $\hat{b}^\dagger$ {and} in the quadrature basis $(\hat{x}, \hat{p})$. The time evolution of the displacement vector in the $(\hat{b}, \hat{b}^\dagger)$ basis is governed by Equation \eqref{bt} and its complex conjugate.

In the quadrature basis $(\hat{x}, \hat{p})$, the displacement vector $\hat{\boldsymbol{O}}(t)$ evolves as: $\hat{\boldsymbol{O}}(t) = S(t) \, \hat{\boldsymbol{O}}(0)$
 where $\hat{\boldsymbol{O}}(t) = \left( \langle \hat{x} \rangle(t), \langle \hat{p} \rangle(t) \right)^\mathrm{T}$, and $S(t) = e^{-M t}
$ is the symplectic evolution matrix with the drift matrix \begin{equation}
M = 
\begin{pmatrix}
\frac{\gamma}{2} & -\omega_m \\
\omega_m & \frac{\gamma}{2}
\end{pmatrix}.
\end{equation}
The matrix $S(t)$ describes a damped rotation in phase space
\begin{equation}\label{Smatrix}
S(t) =
e^{-\frac{\gamma}{2}t}
\begin{pmatrix}
\cos(\omega_m t) & \sin(\omega_m t) \\
-\sin(\omega_m t) & \cos(\omega_m t)
\end{pmatrix}.
\end{equation} This leads to an exponential decay of the displacement amplitude at a rate $\gamma/2$, combined with coherent oscillations at the mechanical frequency $\omega_m$. In the bosonic operator basis $(\hat{b}, \hat{b}^\dagger)$, the covariance matrix $C(t)$ evolves as 
\begin{equation}
C(t) = D(t) \, C(0) \, D^\dagger(t) + C_\mathrm{noise}(t),
\end{equation}
where
\begin{equation*}
    D(t) = 
    \begin{pmatrix}
    e^{-(i\omega_m + \frac{\gamma}{2})t} & 0 \\
    0 & e^{(i\omega_m - \frac{\gamma}{2})t}
    \end{pmatrix},
\end{equation*}
and $C_\mathrm{noise}(t)$ is the noise-induced covariance matrix accumulated during the evolution due to both thermal and gravitational decoherence.

\begin{itemize}
    \item $C(0)$ is the covariance matrix of the initial state in the $\{\hat{b}, \hat{b}^\dagger\}$ basis,
    \item $D(t) = 
    \begin{pmatrix}
    e^{-(i\omega_m + \frac{\gamma}{2})t} & 0 \\
    0 & e^{(i\omega_m - \frac{\gamma}{2})t}
    \end{pmatrix}
    $
    is the evolution matrix in this basis,
    \item $C_\mathrm{noise}(t)$ is the noise-induced covariance matrix accumulated during the evolution due to both thermal and gravitational decoherence.
    \item all elements of $C(t)$ are explicitly given in Equations \eqref{bt}-\eqref{b2t}.
\end{itemize}




In the quadrature basis, the general solution for the covariance matrix is given by \cite{adesso2007entanglementThesis}
\begin{equation}\label{eqcov}
\boldsymbol{\sigma}(t) = S(t) \, \boldsymbol{\sigma}(0) \, S^\mathrm{T}(t) + \boldsymbol{\sigma}_\mathrm{noise}(t).
\end{equation}
Here, $\boldsymbol{\sigma}(0)$ is the initial covariance matrix in the $(\hat{x}, \hat{p})$ basis, $S(t)$ is the symplectic evolution matrix from Eq.~(\ref{Smatrix}), and $\boldsymbol{\sigma}_\mathrm{noise}(t)$ is the particular solution associated with the noise. This structure clearly separates the coherent evolution of the initial state from the contributions due to thermal and gravitational noise.

The full covariance matrix, in the $(\hat{x}, \hat{p})$ quadrature basis, can thus be written explicitly as
\begin{equation}
\boldsymbol{\sigma}(t)
= e^{-\gamma t}\,
\mathbf{R}(\omega_m t)\,
\boldsymbol{\sigma}(0)\,
\mathbf{R}^{\mathrm{T}}(\omega_m t)
+ \boldsymbol{\sigma}_{\mathrm{noise}}(t),
\end{equation} with \begin{equation}
\mathbf{R}(\theta)=
\begin{pmatrix}
\cos\theta & \sin\theta\\
-\sin\theta & \cos\theta
\end{pmatrix}, 
\end{equation} and \begin{equation}
\boldsymbol{\sigma}_{\mathrm{noise}}(t)=
\begin{pmatrix}
N(t)+A(t) & B(t)\\
B(t) & N(t)-A(t)
\end{pmatrix}. \end{equation} The parameters are given explicitly by \begin{eqnarray}
N(t) &=& \left(1 - e^{-\gamma t}\right) (\bar{n}_\mathrm{eff}+\frac{1}{2}), \\
A(t) &=& \frac{2\Lambda}{\gamma^2 + 4\omega_m^2}
\Big[
\gamma \left(1 - e^{-\gamma t} \cos(2\omega_m t)\right) \nonumber \\ && + 
2\omega_m e^{-\gamma t} \sin(2\omega_m t)
\Big], \\
B(t) &=& \frac{2\Lambda}{\gamma^2 + 4\omega_m^2}
\Big[
-2\omega_m \left(1 - e^{-\gamma t} \cos(2\omega_m t)\right)  \nonumber \\ && +
\gamma e^{-\gamma t} \sin(2\omega_m t)
\Big].
\end{eqnarray}
One can see how the covariance matrix evolves under the combined action of mechanical dissipation, coherent oscillations, and both thermal and gravitational decoherence. The first term represents the damped coherent evolution of the initial covariance matrix $\boldsymbol{\sigma}(0)$, while the second term accounts for the accumulation of noise due to diffusion. In the long-time limit ($t \to \infty$), the homogeneous term $S(t) \boldsymbol{\sigma}(0) S^\mathrm{T}(t)$ vanishes exponentially, and the system reaches the stationary state determined solely by $\sigma_\mathrm{noise}(\infty)$, independent of the initial conditions.

\subsubsection*{Particular Solution for the Covariance Matrix}

{In this section we consider the case in which the homogeneous part can be dropped, i.e.,  when the system is initialised in a state with zero displacement and no squeezing. Note that the stationary state (obtained for $t\rightarrow +\infty$) falls in this class and, in particular, depends only on the parameters $\omega_m,\gamma, \Lambda_g, \Lambda_T$ as the homogeneous part exponentially decays. Within these conditions, the covariance matrix 
in the case $t\rightarrow +\infty$, becomes}
\begin{equation}
\sigma(\infty) =
\begin{pmatrix}
\bar{n}_{\mathrm{eff}} + \frac{1}{2}  + A_\infty & B_\infty \\
B_\infty & \bar{n}_{\mathrm{eff}} + \frac{1}{2} - A_\infty
\end{pmatrix},
\end{equation}
where $A_\infty$ and $B_\infty$ encode quadrature squeezing and  correlations between position and momentum, respectively.

This structure reveals that the steady state is not a simple thermal state, but rather a \textit{squeezed thermal state} with non-zero correlations between quadratures. The decoherence processes, including the gravitational contribution, leave signatures not only as isotropic noise (heating) but also as ``deformation'' and inclination of the uncertainty ellipse in phase space. More explicitly, the squeezing term $A_\infty$ and the $xp-$correlation term $B_\infty$ are
\begin{align}
A_\infty &= \frac{2\Lambda \gamma}{\gamma^2 + 4\omega_m^2}, ~~
B_\infty = -\frac{4\Lambda \omega_m}{\gamma^2 + 4\omega_m^2}.
\end{align}

Both $A_\infty$ and $B_\infty$ scale linearly with $\Lambda$, {thus with} $\Lambda_g$. This indicates that gravitational decoherence directly affects the size, shape, and orientation of the Wigner function in phase space. {Also, note that their ratio can be expressed solely in terms of the mechanical heat $Q=\omega_m/\gamma$, that is
\begin{equation}
    \frac{B_\infty}{A_\infty}= -2Q.
\end{equation}}

The denominators $\gamma^2 + 4\omega_m^2$ appearing in both expressions highlight the balance between dissipation ($\gamma$) and coherent oscillations ($\omega_m$). Specifically: (i) In the {overdamped regime} ($\gamma \gg \omega_m$), the expressions simplify to $A_\infty \approx \frac{2\Lambda}{\gamma}$ and $B_\infty \approx 0$; the squeezing vanishes, and the steady state approaches a simple thermal state with isotropic noise. (ii) In the {underdamped regime} ($\omega_m \gg \gamma$), one obtains $A_\infty \approx 0$ and $B_\infty \approx -\frac{\Lambda}{\omega_m}$; in this case, the system exhibits strong quadrature correlations, a direct consequence of the oscillatory dynamics of the mode.

The fact that $\Lambda_g$ explicitly appears in all terms of the covariance matrix has direct implications for quantum parameter estimation. Gravitational decoherence not only increases the thermal occupation of the mode (through $\bar{n}_{\mathrm{eff}}$), but also induces characteristic squeezing and cross-quadrature correlations. These signatures are formally indistinguishable from those arising from thermal decoherence in terms of their mathematical structure in the master equation. However, their dependence on $\Lambda_g$ is explicit and measurable: from a quantum metrology perspective, the QFI quantifies exactly how sensitive the steady state is to changes in $\Lambda_g$. This sensitivity arises both from the enlargement of the uncertainty (through increased noise) and from its ``deformation'' in phase space (through squeezing for example).

One can conclude from the stationary analysis of the covariance matrix, that gravitational decoherence manifests as an additional diffusion process that not only heats the mechanical mode but also induces squeezing and correlations between position and momentum quadratures. These features are encoded directly in the covariance matrix and provide the necessary structure to enable precision estimation of $\Lambda_g$. Also, note that in the stationary regime, the covariance matrix becomes independent of the initial state. This is a general property of Gaussian systems subject to Markovian dissipation and diffusion, where the dynamics ``erase'' any initial squeezing, displacement, or thermal occupation, leading to a steady state fully determined by the system parameters $(\omega_m, \gamma)$ and the decoherence rates $(\Lambda_T, \Lambda_g)$. Consequently, also the QFI in the stationary limit depends solely on these parameters and not on the initial conditions.

We would like to stress that, from a physical point of view, the under-damped regime ($\omega_m \gg \gamma$) is more consistent with realistic optomechanical systems, particularly those inspired by gravitational-wave detectors. These setups are designed to preserve coherence over many oscillation cycles, making the condition $\omega_m \gg \gamma$ not only analytically convenient but also experimentally relevant.

\subsection{Quantum Fisher Information for a single-mode Gaussian state}

The QFI ${\mathcal F}_{Q}$ quantifies the ultimate precision bound for estimating the gravitational decoherence parameter $\Lambda_g$ encoded in the Gaussian state of the mechanical mode. For a general single-mode Gaussian state, characterized by the covariance matrix $\boldsymbol{\sigma}(t)$ and the displacement vector $\hat{\boldsymbol{O}},$ the QFI with respect to $\Lambda_g$ is given by~\cite{Pinel2013, vsafranek2018simple, monras2013phase, vsafranek2018estimation}

\begin{widetext}
\begin{equation}
{\mathcal F}_{{Q}}(t, \Lambda_g) =
\frac{1}{2(1 + P(t)^2)} 
\mathrm{Tr}\left[
\left( \sigma^{-1}(t) \, \frac{\partial \boldsymbol{\sigma}(t)}{\partial \Lambda_g} \right)^2
\right]
+
\frac{2}{1 - P(t)^4}
\left(
\frac{\partial P(t)}{\partial \Lambda_g}
\right)^2
+
\left(
\frac{\partial \hat{\boldsymbol{O}}^\mathrm{T}(t)}{\partial \Lambda_g}
\right)
\sigma^{-1}(t)
\left(
\frac{\partial \hat{\boldsymbol{O}}(t)}{\partial \Lambda_g}
\right) \label{pinel}
\end{equation}
\end{widetext}
with the purity $P(t)$ defined as \cite{adesso2007entanglementThesis, adesso2005purity, cerf2007quantum,  ferraro2005gaussian, souza2008characteristic, souza2008quantifying}
\begin{equation}
P(t) = \frac{1}{2\sqrt{\det \boldsymbol{\sigma}(t)}}. \label{purity}
\end{equation}
The first two terms correspond to the sensitivity of the covariance matrix --- capturing both the shape (squeezing and correlations) and the global mixedness (via purity). The third term arises from the sensitivity of the displacement vector $\hat{\boldsymbol{O}}(t) = \left( \langle \hat{x} \rangle(t), \langle \hat{p} \rangle(t) \right)^\mathrm{T}$ with respect to $\Lambda_g$. This contribution vanishes when the state has no displacement, but is present in general for coherent or displaced Gaussian states.

In the stationary limit ($t \to \infty$), the homogeneous evolution of the displacement decays exponentially due to mechanical damping, and the contribution from the displacement term vanishes. The QFI then reduces to the contribution arising purely from the covariance matrix. In contrast, for finite times or systems prepared in displaced Gaussian states, the displacement term can significantly contribute to the QFI.

This structure highlights that gravitational decoherence influences the estimation precision through two distinct mechanisms: (i) by modifying the covariance matrix, which affects both the uncertainty volume (via purity) and its shape (via squeezing and quadrature correlations); and (ii) by affecting the evolution of the displacement vector when present. Both mechanisms contribute to the total sensitivity to $\Lambda_g$, and consequently to the fundamental quantum limits on precision in any metrological protocol based on mechanical Gaussian states.

\subsection{Estimation Strategy}


Our goal is to estimate the gravitational decoherence parameter $\Lambda_g$ by exploiting quantum features of optomechanical systems inspired by gravitational-wave detectors. The strategy we propose combines tools from quantum estimation theory with experimentally feasible measurement protocols, as outlined below.

First, we consider the preparation of a single-mode Gaussian probe state. In line with techniques already employed in gravitational-wave interferometry, we investigate several classes of input states, including coherent states, thermal states, squeezed vacuum states, and thermal squeezed states. These probes evolve under a dissipative dynamics influenced by thermal noise and gravitational decoherence. The system's evolution is captured through a time-dependent covariance matrix $\boldsymbol{\sigma}(t)$ and mean vector $\hat{\boldsymbol{O}}(t)$, which encode the effects of the environment and the gravitational parameter.

The quantum parameter $\Lambda_g$ is imprinted onto the state during this evolution. To extract information about it, we perform a homodyne measurement of a quadrature observable, either $\hat{x}$ or $\hat{p}$, which results in a Gaussian probability distribution whose width depends on $\Lambda_g$. This approach has the advantage of being physically implementable in state-of-the-art optomechanical platforms \cite{bose2025massive}.

To evaluate the ultimate estimation precision, we calculate the QFI associated with the evolved state. 
It is computed as a function of the evolution time $t$ and the input state parameters. By analyzing the QFI, one can in principle identify the optimal time $t_{\text{opt}}$ that maximizes the sensitivity to $\Lambda_g$.

Repeating the experiment $N$ times allows us to approach the quantum Cramér-Rao bound.
This estimation strategy employs realistic measurement schemes, takes advantage of squeezing and quantum statistical effects, and aligns with recent advances in quantum-enhanced metrology \cite{ligomemorandum, aasi2013enhanced, ganapathy2023broadband, bose2025massive}.

\section{QFI for different single-mode Gaussian states}

{Now we proceed to analise QFI for different Gaussian states. Since we study single-mode Gaussian states, properties like entanglement cannot be used as a classifying criterion (see for example \cite{Leo2024}). Single-mode states have extensively been characterised in \cite{adesso2007entanglementThesis, adesso2007entanglement, ferraro2005gaussian, souza2008characteristic, souza2008quantifying}.

We study the numerical solution of Eq.~\eqref{pinel} with different initial conditions but with fixed energy $n_0=4$.} {We are interested in the following cases}: coherent states ({of} displacement $\alpha_0$), squeezed vacuum states ({of} initial squeezing $r$), thermal states (with an initial thermal excitations $n_{\text{th}0}$), and squeezed thermal states (with initial squeezing $r$ and thermal excitations  $n_{\text{th}0}$). Initial states and parameters used in the numerical analysis are summed up in Table~\ref{tab:initial_states}. 


%
\begin{table}[t!]
\caption{Initial states and parameters in the simulations.}
\label{tab:initial_states}
\begin{ruledtabular}
\begin{tabular}{cc} 
\hline
\textbf{Initial State} & \textbf{Parameters}                  \\
\hline
Coherent & $|\alpha_0| = 2$ \\
Thermal & $n_{\text{th}0} = 4$ \\
Squeezed vacuum & $r = 1.4436$ \\
Squeezed thermal & \makecell[l]{$\left\{\begin{aligned}
&n_{\mathrm{th}0} = 1 \\
&r = 0.8814
\end{aligned}\right.$} \\ 
\hline
\end{tabular}
\end{ruledtabular}
\end{table}

\begin{figure}[t!]
    \centering
    \includegraphics[width=0.9\columnwidth]{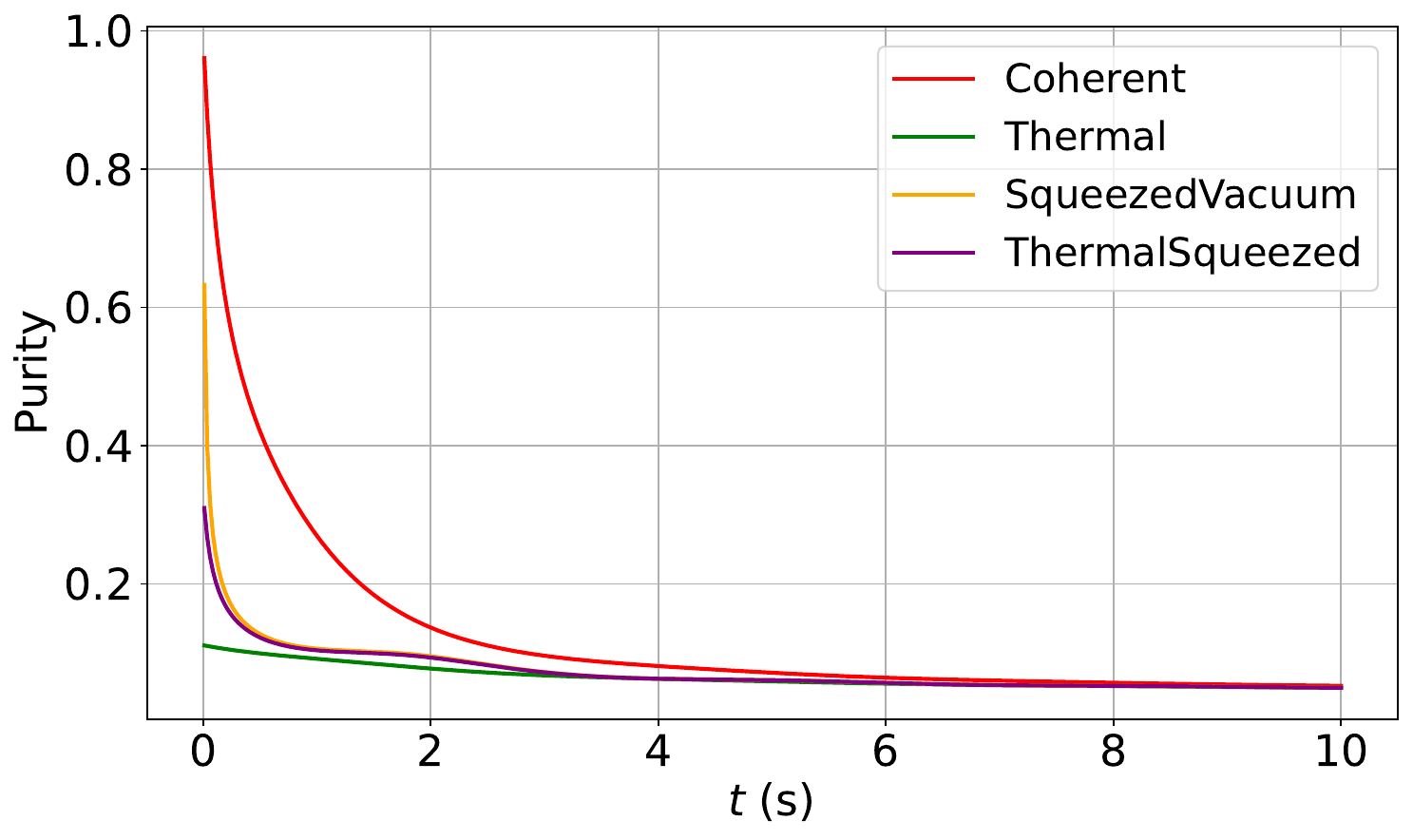}
    \caption{{Evolution of purity for initial energy $n_0=4$ in different quantum states. As expected, purity decreases with time due to dissipation and decoherence. Squeezed states exhibit faster decoherence compared to coherent states, indicating greater susceptibility to environmental interactions.}}
    \label{fig:purity}
\end{figure}

\begin{figure}[t!]
    \centering
    \includegraphics[width=0.9\linewidth]{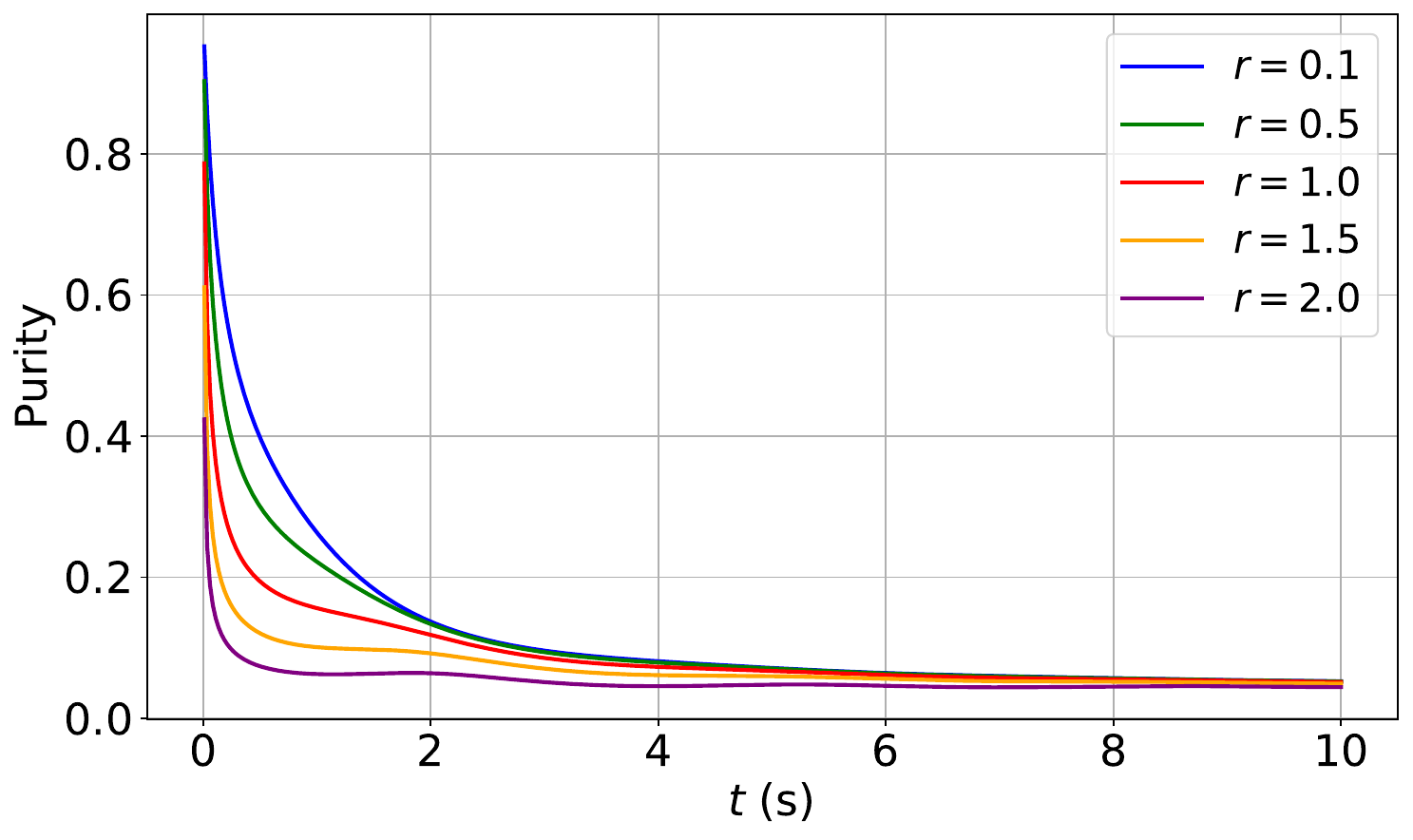}
    \vspace{0.5em}
    \includegraphics[width=0.9\linewidth]{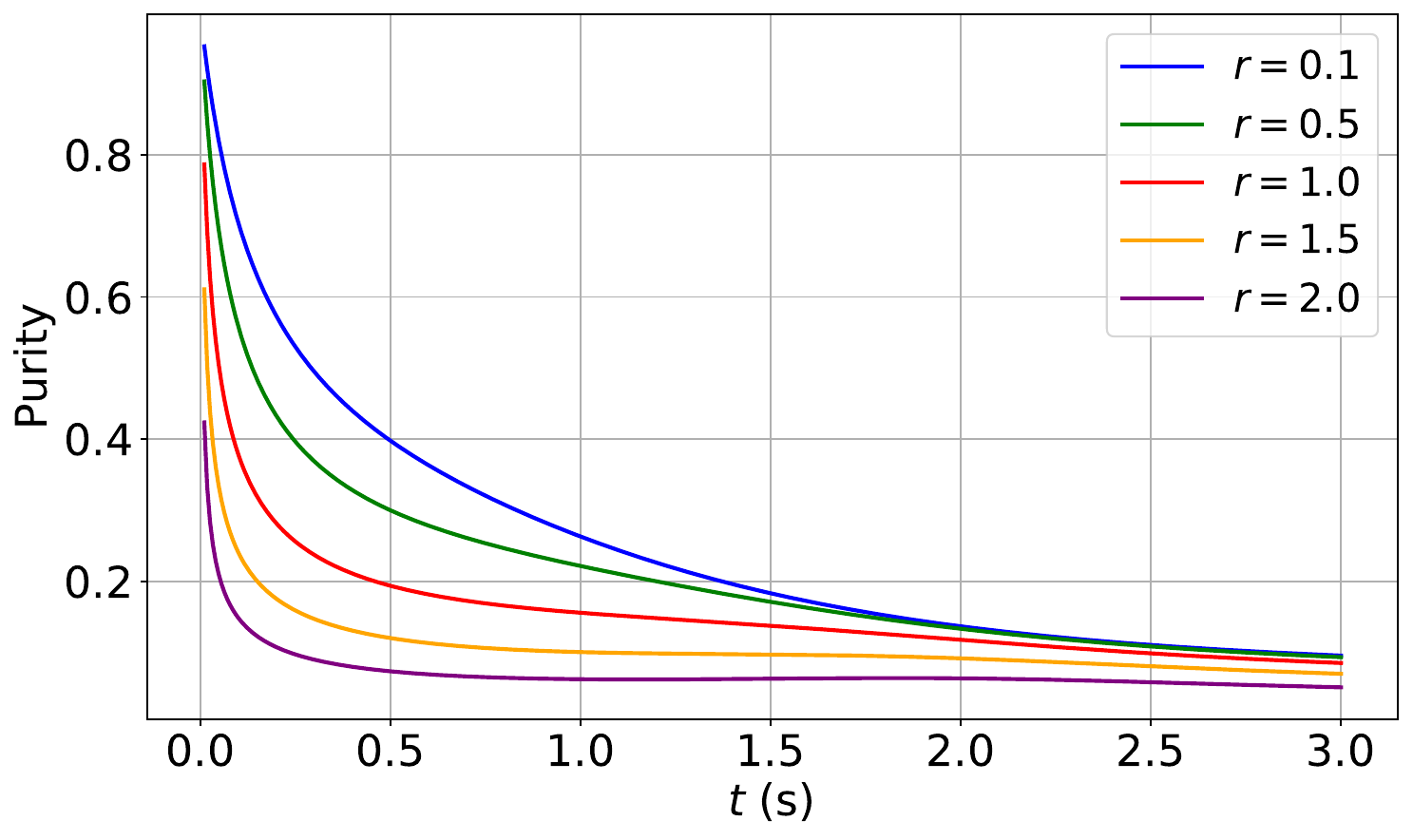}
    \caption{Purity of single-mode squeezed vacuum states for various squeezing parameters $r$, with $\Lambda_g = 10^{-8}$. Top: long-time behavior. Bottom: transient dynamics with short-time oscillations due to the interplay between gravitational decoherence and thermal noise. States with higher squeezing show faster purity decay, indicating greater sensitivity to dissipation.}
    \label{fig:purityr}
\end{figure}

Since the calculation of the QFI requires the evaluation of the state purity given in Eq.~\eqref{purity}, we also present the purity for {the different} states. This quantity is well studied in the literature; for a detailed discussion on the purity of Gaussian states, we refer the reader to \cite{adesso2005purity, souza2008characteristic, souza2008quantifying}. In Fig.~\ref{fig:purity} we show how purity evolves in time, while Fig.~\ref{fig:purityr} displays the same evolution under different initial squeezing values for squeezed vacuum states.

\begin{figure}[!]
    \centering
    \includegraphics[width=1\columnwidth]{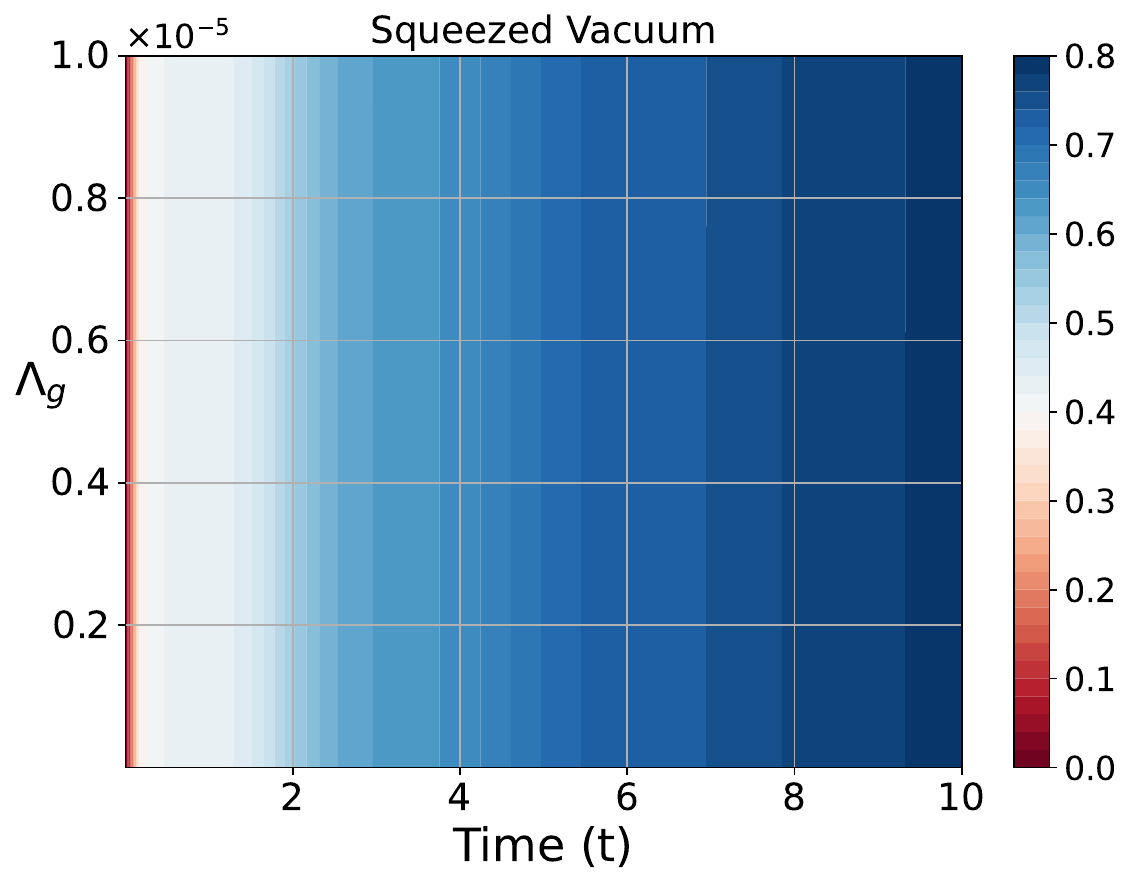}
    \caption{QFI as a function of $t$ and $\Lambda_g$, with squeezed vacuum as initial state. One can see clearly that the QFI asymptotically approaches a steady value, as decoherence acts.}
    \label{fig:contour_QFI}
\end{figure}

\begin{figure}[t!]
    \centering
    \includegraphics[width=1\columnwidth]{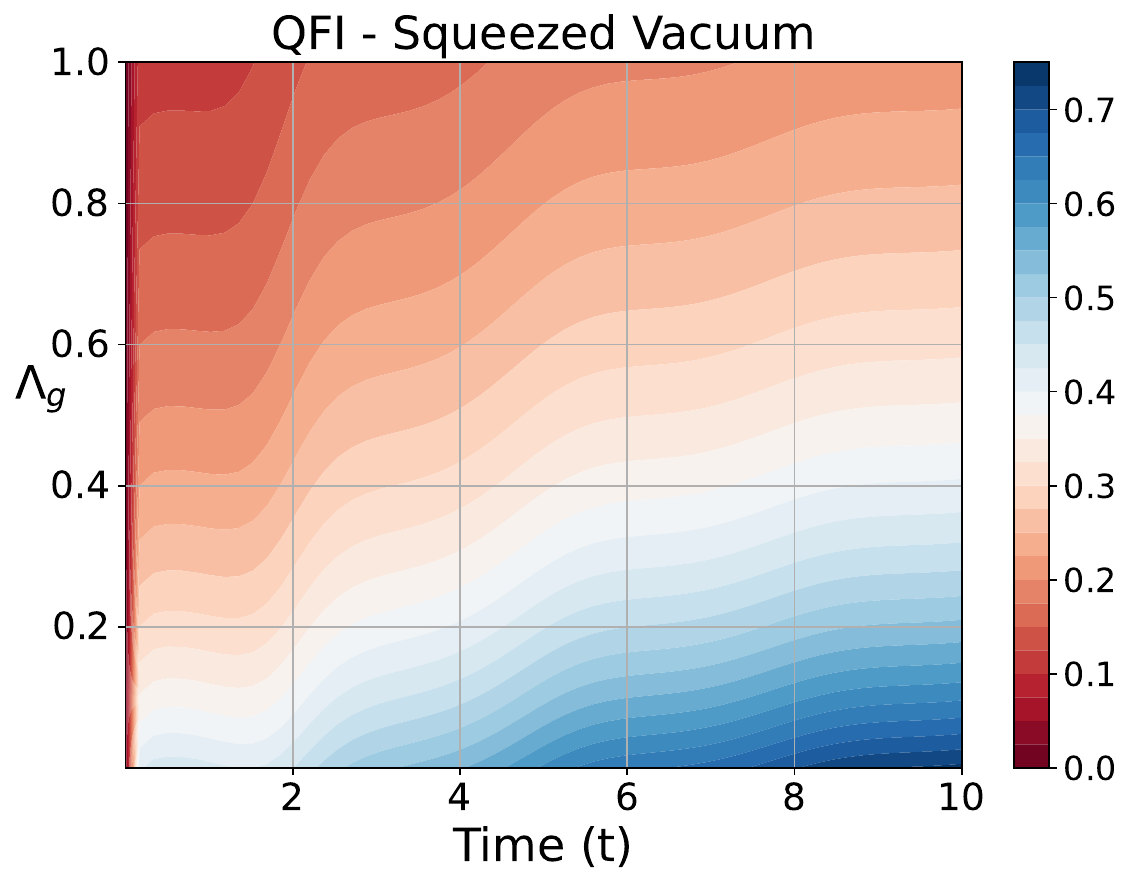}
    \caption{QFI as a function of time $t$ and the parameter $\Lambda_g$, using squeezed vacuum states as probe. For large (unphysical) values of $\Lambda_g$, the QFI drops rapidly, highlighting the protocol’s sensitivity within the physically relevant regime.}
    \label{fig:contour_QFI1}
\end{figure}

\begin{figure}[t!]
    \centering
    \includegraphics[width=1\linewidth]{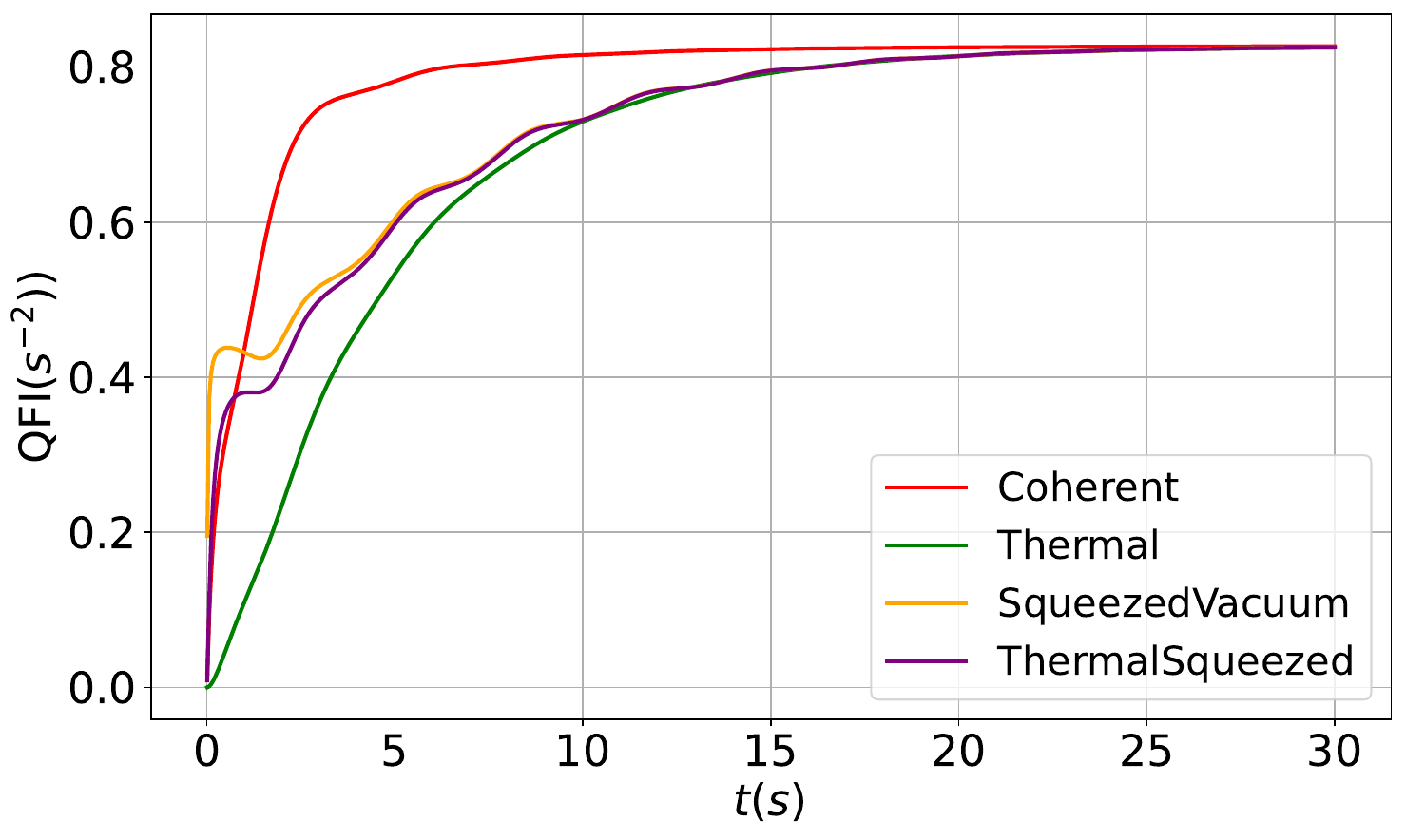}
    \vspace{0.5em}
    \includegraphics[width=1\linewidth]{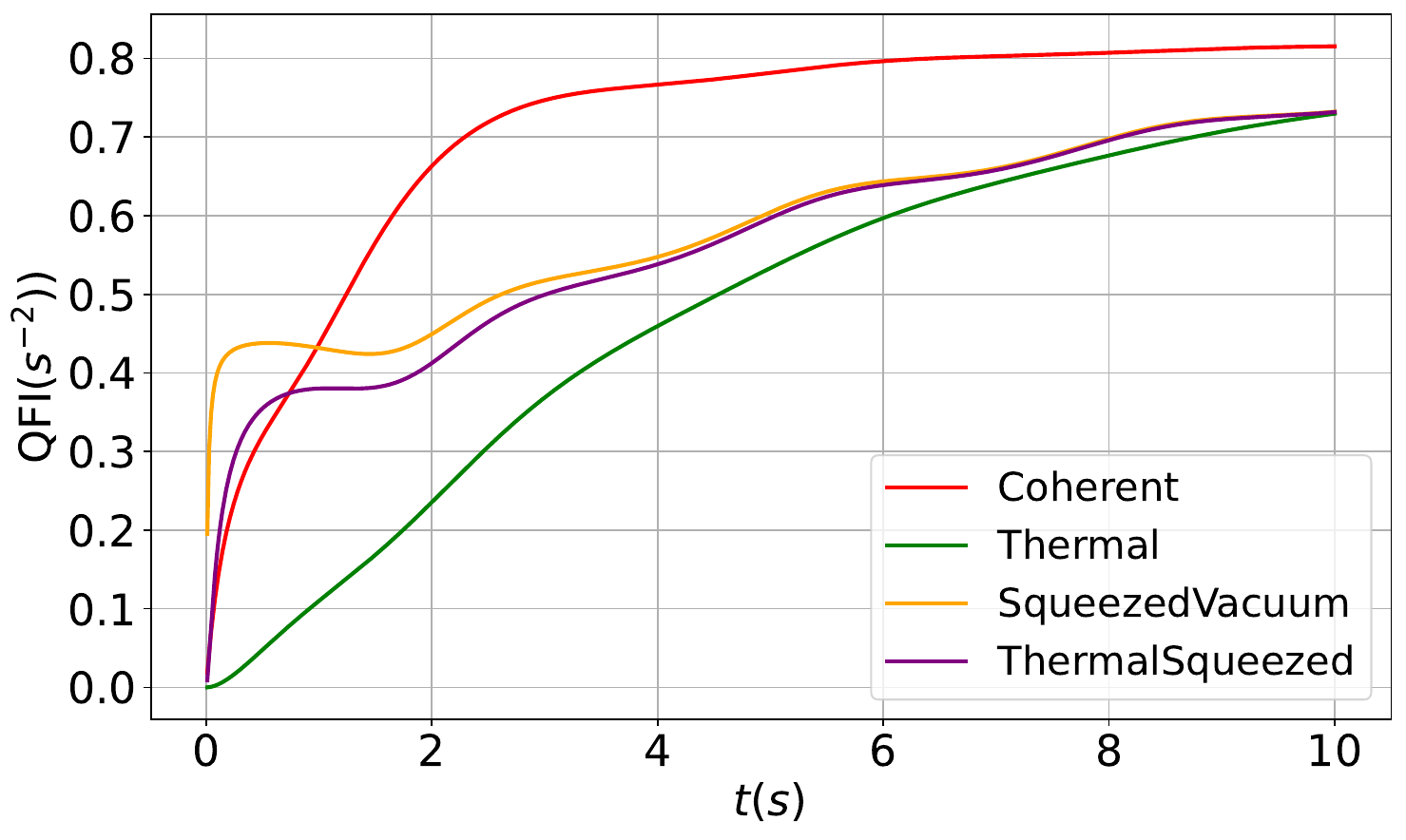}
    \caption{QFI as a function of $t$ for the different classes of single-mode Gaussian states. Top: long-time behaviour. Bottom: transient dynamics with short-time oscillations due to the interplay between gravitational decoherence and thermal noise. States with higher squeezing show faster purity decay, indicating greater sensitivity to dissipation.}
    \label{fig:QFI_classes}
\end{figure}

\begin{figure}[t!]
    \centering    \includegraphics[width=1\linewidth]{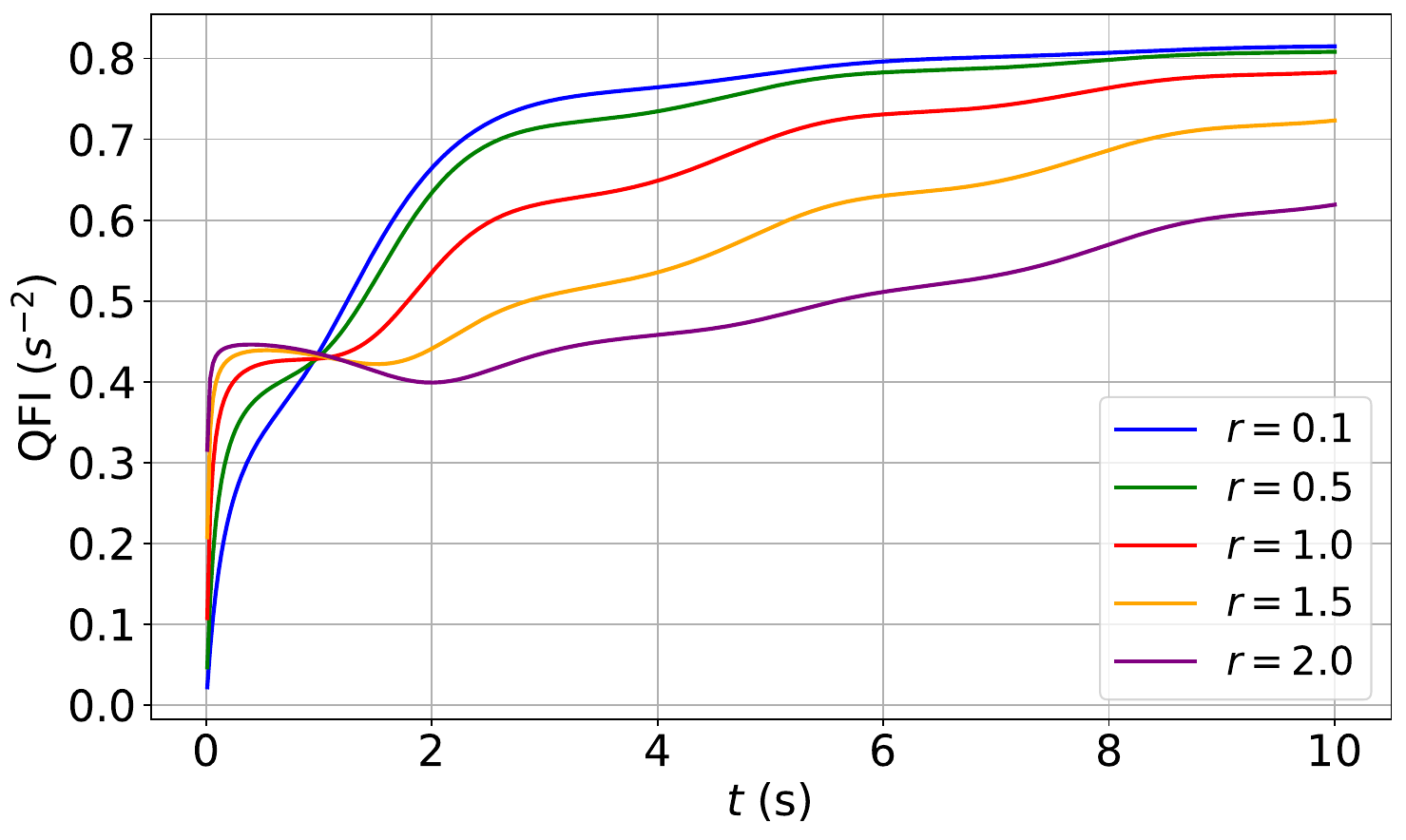}
    \vspace{0.5em}
    \includegraphics[width=1\linewidth]{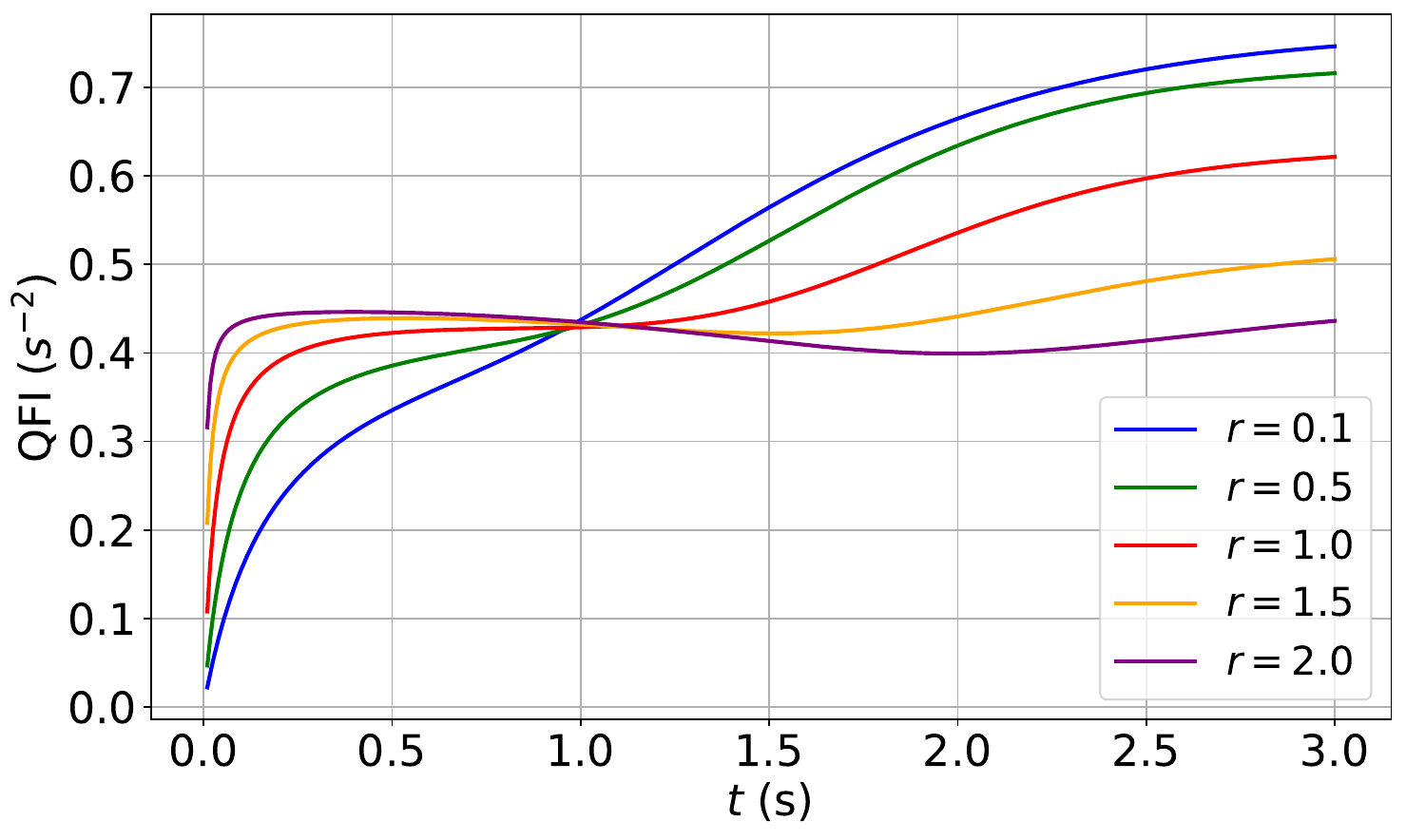}
    \caption{QFI as a function of time $t$ for squeezed vacuum states used as probes. Top: long-time behavior. Bottom: transient dynamics. Higher initial squeezing $r$ yields greater QFI at short times. Due to stronger decoherence associated with larger squeezing, probes with smaller $r$ eventually surpass the more squeezed ones at longer times.}
    \label{fig:QFI_squeezed}
\end{figure}

{Next, we evaluate the QFI of Eq.~\eqref{pinel} to bound the estimation precision of the gravitational decoherence rate $\Lambda_g$. The results are shown in Figs.~\ref{fig:contour_QFI}-\ref{fig:QFI_squeezed}.} {We observe that the qualitative behaviour of QFI in short-time is almost constant with respect to small changes in $\Lambda_g$, in particular in the experimentally expected range. After a transient it approaches an asymptotic value, as predicted in the sections above. For this reason, we make an educated guess of the experimental value of $\Lambda_g$ to address the question of which initial state is better suited to detect its values within the window of time allowed by experiments. }

In Fig.~\ref{fig:contour_QFI}, we show the behaviour of the QFI as a function of time $t$ and the gravitational decoherence parameter $\Lambda_g$, using the squeezed vacuum state as probe. We observe that this qualitative behaviour is similar across all classes of states (though with distinct QFI values for each). 

In Fig.~\ref{fig:contour_QFI1}, we present the QFI as a function of both $t$ and $\Lambda_g$ over an extended parameter region, for squeezed vacuum as initial state. This plot shows that, for unrealistically large values of $\Lambda_g$, the QFI drops sharply, reinforcing the sensitivity of the estimation protocol to this parameter within physically meaningful regimes.

{Since we expect $\Lambda_g$ to be of the order of $10^{-8}\ \mathrm{s}^{-1}$ for a suspended mirror with mechanical frequency $\omega_m$ of the order of $\mathrm{s}^{-1}$, we fix this value and show in Fig.~\ref{fig:QFI_classes} that squeezed vacuum states outperform the others at short times.} This behaviour can be attributed to their higher degree of ``quantumness'' (as reflected, for instance, in their second-order correlation function $g^{(2)}(0)$ or in the dynamics of their mean excitation number \cite{adesso2005purity, souza2008characteristic, souza2008quantifying}). However, beyond this short-time regime, squeezed vacuum states decohere more rapidly, and coherent states eventually may surpass them as more robust probes for estimating $\Lambda_g$.

{In Fig.~\ref{fig:QFI_squeezed} we characterise squeezed vacuum states by different squeezing parameters, confirming the sensitivity of these states to decoherence effects\cite{adesso2007entanglementThesis, adesso2007entanglement, ferraro2005gaussian, souza2008characteristic, souza2008quantifying}}. As a result, the transient advantage provided by squeezing is quickly lost, and for longer times, states with lower initial squeezing ultimately outperform highly squeezed states in terms of QFI.

\section{Conclusions}

In this work we have developed and numerically validated a quantum estimation framework to assess the precision with which a gravitationally induced decoherence rate $\Lambda_g$ can be inferred from the dynamics of a mechanically compliant mode, modeled within a Gaussian open-system description inspired by gravitational-wave interferometry. By expressing the evolution in terms of the time-dependent covariance matrix and by employing the general single-mode Gaussian QFI formula \cite{Pinel2013}, we have identified how $\Lambda_g$ leaves measurable signatures.

Our central metrological result is that, squeezed vacuum probes provide the largest QFI at short times, outperforming coherent, thermal, and squeezed-thermal states in the experimentally relevant regime where $\Lambda_g$ is expected. This transient enhancement is physically intuitive: squeezing amplifies the susceptibility of the covariance matrix to diffusion-induced dynamics, thereby increasing the distinguishability of nearby values of $\Lambda_g$. However, we also find a clear trade-off between quantumness and robustness: the nonclassical features that boost the short-time sensitivity also accelerate the loss of purity, so that the advantage of highly squeezed probes is progressively eroded by dissipation or noise. As a consequence, for longer evolution times the QFI of coherent probes can match and even surpass that of squeezed vacuum states, highlighting that optimal estimation is not dictated by squeezing alone, but by the balance between information gain and decoherence-induced degradation.

These findings have direct implications for experimental design. First, they indicate that protocols targeting gravitational decoherence should operate in the underdamped regime ($\omega_m\gg\gamma$), which is consistent with optomechanical platforms motivated by gravitational-wave detection; and also should prioritize short-time, where squeezing-induced sensitivity is maximal. Second, they provide a concrete, state-dependent benchmark for realistic measurement schemes (e.g., homodyne detection of a mechanical quadrature), enabling a selection of probe preparation parameters and optimal measurement times. Finally, our analysis establishes a quantitative bridge between proposed models of gravitational decoherence and experimentally accessible metrological figures of merit, offering ways to compare different decoherence mechanisms on the same operational footing.

Several extensions may follow from the present study. On the theoretical side, it may be interesting to optimize over both probe preparation and measurement strategy, and to assess explicitly the gap between the QFI and the classical Fisher information achievable with experimentally constrained observables. On the physical side, incorporating additional ingredients beyond the single-mode Gaussian setting, such as cavity-assisted readout, non-Markovian noise, multi-mode correlations, or genuinely entangled continuous variable probes, may further enhance the sensitivity to $\Lambda_g$ and clarify the role of quantum correlations in fundamental decoherence estimation. Therefore, the framework presented here provides a baseline for these developments and supports the feasibility of using optomechanical quantum metrology as a tool to probe gravitationally induced decoherence in realistic settings \cite{pfister2016universal}.

\section*{Acknowledgments}

L.A.M.S. acknowledges the Instituto Nacional de Ci\^encias e Tecnologia -- Informa\c{c}\~ao Qu\^antica (INCT--IQ). OPSN is grateful to FAPEPI. Partially supported by  FAPEPI, via EDITAL N$^{o}$ 04/2025: {\it ``PROP/UESPI/FAPEPI.''}. E.R. and R.L.F. acknowledge support from the PNRR Project PRISM – Partenariato Esteso RESTART – PE00000001 – Spoke 4 – CUP: E13C22001870001.  G.A. acknowledges support from the UK
Research and Innovation (UKRI) via EPSRC Grant No.~EP/X010929/1.

\bibliography{refs_leo, apssamp}

\end{document}